\documentclass{article}
\usepackage{spconf,amsmath,graphicx}

\usepackage{cite}
\usepackage{bm}
\usepackage{xcolor}
\usepackage{hyperref}
\usepackage{booktabs}
\usepackage{paralist}
\usepackage{float}
\usepackage{caption} 
\usepackage{subcaption}
\usepackage{fancyhdr}
\usepackage{multirow}
\usepackage{comment}
\usepackage{array}
\usepackage{balance}
\newcolumntype{P}[1]{>{\centering\arraybackslash}p{#1}}
\newcolumntype{R}[1]{>{\raggedleft\arraybackslash}p{#1}}
\newcommand{\sstitle}[1]{\smallskip\noindent\textbf{#1.\/}}

\newcommand{\eg}{e.\,g.,\ }
\newcommand{\ie}{i.\,e.,\ }

\ninept

\title{Fast Yet Effective Speech Emotion Recognition with Self-Distillation}

\name{Zhao Ren${^1}$, Thanh Tam Nguyen${^2}$, Yi Chang${^3}$, Björn W. Schuller${^{3,4}}$ \thanks{This research was funded by the Federal Ministry of Education and Research (BMBF), Germany under the project LeibnizKILabor with grant No.\,01DD20003, and the research projects ``IIP-Ecosphere”, granted by the German Federal Ministry for Economics and Climate Action (BMWK) via funding code No.\,01MK20006A.
}}
\address{$^1$L3S Research Center, Leibniz University Hannover, Germany\\
$^2$Griffith University, Australia\\
$^3$GLAM – Group on Language, Audio, \& Music, Imperial College London, United Kingdom\\
$^4$Chair of Embedded Intelligence for Health Care and Wellbeing, University of Augsburg, Germany\\
{\small \tt zren@l3s.de}
}

\begin{document}
%\ninept
%
\maketitle
\begin{abstract}
Speech emotion recognition (SER) is the task of recognising human's emotional states from speech. SER is extremely prevalent in helping dialogue systems to truly understand our emotions and become a trustworthy human conversational partner. Due to the lengthy nature of speech, SER also suffers from the lack of abundant labelled data for powerful models like deep neural networks. Pre-trained complex models on large-scale speech datasets have been successfully applied to SER via transfer learning. However, fine-tuning complex models still requires large memory space and results in low inference efficiency. In this paper, we argue achieving a fast yet effective SER is possible with self-distillation, a method of simultaneously fine-tuning a pretrained model and training shallower versions of itself. The benefits of our self-distillation framework are threefold: (1) the adoption of self-distillation method upon the acoustic modality breaks through the limited ground-truth of speech data, and outperforms the existing models' performance on an SER dataset; (2) executing powerful models at different depth can achieve adaptive accuracy-efficiency trade-offs on resource-limited edge devices; (3) a new fine-tuning process rather than training from scratch for self-distillation leads to faster learning time and the state-of-the-art accuracy on data with small quantities of label information.

\end{abstract}
\section{Introduction}
\label{sec:intro}

Speech emotion recognition (SER) nowadays is an idiosyncratic task in many dialogue systems, such as Siri, Cortana, and Alexa~\cite{schuller2018speech}. Through classifying human speech signals into various emotional states (\eg happiness, surprise, anger, disgust, fear, sadness, neutral, etc.), SER helps human-computer systems become more personalised and trustworthy as well as adjust the contexts accordingly in car-driving, heath-diagnosis, call-center, aircraft-cockpit, and web/mobile applications~\cite{el2011survey,khalil2019speech}.

Existing techniques for SER are limited by the inherent lack of labelled data due to the expensive efforts of annotation (e.g. thousands of hours of speech over nearly 7,000 spoken languages~\cite{baevski2020wav2vec}). They often rely on large deep neural networks that are pre-trained by unsupervised learning, contrastive learning, or self-supervised learning, such as wav2vec~\cite{SchneiderBCA19}, wav2vec 2.0~\cite{baevski2020wav2vec}, and vq-wav2vec~\cite{BaevskiSA20}. 
%However, fine-tuning complex models actually requires large memory space and always results in low inference efficiency~\cite{chang2022distilhubert}.
However, fine-tuning large models has a high demand of memory space and inference time~\cite{chang2022distilhubert}.
%BS: this sentence above is an almost 1:1 copy from the introduction - please check and perhaps rephrase :)
%, while only improving the accuracy marginally~\cite{chang2022distilhubert}.

%\todo{TODO: why performing self-supervision for SER is not enough or non-trivial?}

In machine learning, self-distillation has emerged as a paradigm to develop a student model with a more lightweight architecture that can even outperform the teacher~\cite{zhang2019your}. This has been particularly successfully applied to computer vision~\cite{zhang2021self,zhang2019your}. However, in contrast to the visual modality, the acoustic modality is significantly more challenging due to limited ground-truth. Self-distillation methods cannot be applied directly to SER since they often require large labelled data to simultaneously train a teacher model from scratch with shallower student versions of itself~\cite{zhang2019your}.

In this paper, we present a framework of self-distillation for fast, yet effective speech emotion recognition. While our framework is demonstrated on wav2vec 2.0~\cite{baevski2020wav2vec} (one of the state-of-the-art (SOTA) pre-trained models for speech representations), it can be applied to other models and datasets with limited ground-truth information. 
%\autoref{fig:framework} illustrates the main flow and components of our framework. 
In our framework (see \autoref{fig:framework}), the pre-trained wav2vec 2.0 (\ie the teacher model) was fine-tuned together with shallower model parameters from itself (\ie the student models), when the teacher and all students are predicting emotional states from speech samples.  
%\todo{TODO: briefly describe the framework here and the benefit of each component}.

To the best of our knowledge, this is the first attempt to develop a self-distillation framework for SER. The contributions of our self-distillation framework include: (1) the application of self-distillation on speech data overcomes the difficulty caused by limited annotations, and outperforms the existing models' performance on an SER dataset; (2) executing powerful models at different depths increases the possibility to achieve adaptive accuracy-efficiency trade-offs on resource-limited edge devices; (3) a new fine-tuning process rather than training from scratch for self-distillation leads to faster learning time and SOTA accuracy on data with limited ground-truth.

%Due to the strong capability of wav2vec 2.0 for learning representations, model fine-tuning with the above model structure is promising to outperform models trained from scratch in an SER task.

\sstitle{Related Works}
%\todo{TODO: Briefly mention existing deep learning models for SER and their mechanisms. Emphasize why they either do not have high accuracy or are not lightweight.}
Spectrum features~\cite{gong2022ssast,singh2021multimodal} have been often used as the input of deep neural networks for SER~\cite{wani2021comprehensive}, while selecting the appropriate spectrum features is a time-consuming work. Moreover, the performance of SER is limited to expensive human annotations; lacking of labelled data for deep learning. More recently, self-supervised learning on speech data has shown promising to learn effective representations, and the pre-trained models have been successfully fine-tuned for SER tasks~\cite{pepino2021emotion,chen2021exploring,wagner2022dawn}. Therefore, we apply an end-to-end self-supervised learning model, wav2vec 2.0, to SER.

Knowledge distillation is one of the popular methods to achieve high efficiency by transferring knowledge from a teacher model to a smaller student model~\cite{chang2022distilhubert}. Similar to other model compression approaches such as pruning and quantisation, they sacrifice information loss (thus accuracy) and could not overcome the accuracy-efficiency trade-offs. Our self-distillation approach can achieve the best of both worlds by reusing the architecture and allowing inference at different depths of the teacher model itself.

Different types of self-distillation methods have been developed recently, including iteration-based~\cite{mobahi2020self,pham2022revisiting}, aggregate-based~\cite{lee2020self}, and branch-based approaches~\cite{zhang2019your,zhang2021self}. Iteration-based methods perform knowledge distillation from a teacher model to a student model with the same architecture and this procedure is repeated a few times~\cite{mobahi2020self,pham2022revisiting}. However, the training and inference costs are not reduced, as the teacher and the student are the same. Aggregated-based methods use data augmentation to produce more versions of the teacher model on different augmentations and then combine the outputs~\cite{lee2020self}. However, existing data augmentations are domain-specific and are not applicable to acoustic modality. Our work relates closely to the branch-based approaches, which add branches at different depths of the teacher model using bottlenecks/attention-blocks and shallow classifiers~\cite{zhang2019your,zhang2021self}. However, these layers are not applicable for wave2vec 2.0, as it already contains transformer layers. 
Our work is also different from the layer-wise knowledge distillation, which fine-tunes a student model from the teacher model itself for predicting deep layers of the teacher~\cite{chang2022distilhubert}. The layer-wise knowledge distillation can produce a general student model, while it requires fine-tuning efforts for a specific task. The fixed number of model parameters of the student is limited for performance improvement with a deeper structure and lacks of flexibility.

%\todo{TODO: explain our design choices compared to others more} 

\section{Methodology}
\label{sec:method}

\subsection{Preliminaries}
\label{sec:wav2vec}

\sstitle{Self-supervised Learning with wav2vec 2.0}
Self-supervised learning has shown its superiority compared to supervised learning on many audio tasks, such as speaker recognition~\cite{vaessen2022fine} and SER~\cite{mohamed2021arabic}. Wav2vec 2.0~\cite{baevski2020wav2vec} was trained on the large-scale Librispeech corpus~\cite{panayotov2015librispeech} in a self-supervised learning framework. Wav2vec 2.0 has been widely used to extract effective representations for SER tasks~\cite{pepino2021emotion,wongpatikaseree2022real}.
A Wav2vec 2.0 model consist of multi-layer convolutional neural networks (CNNs) (\ie encoder) and multiple transformer layers (\ie context network). The latent representations learnt from the encoder are discretised into a set of quatisation representations, which are then processed with the output of the context network in a contrastive task.  

\subsection{Self-Distillation Framework: The Case of Wav2vec 2.0}
\label{sec:selfdistill}

\subsubsection{Model Architecture}
\begin{figure}
    \centering
    \includegraphics[width=\linewidth,trim={120 50 20 80},clip]{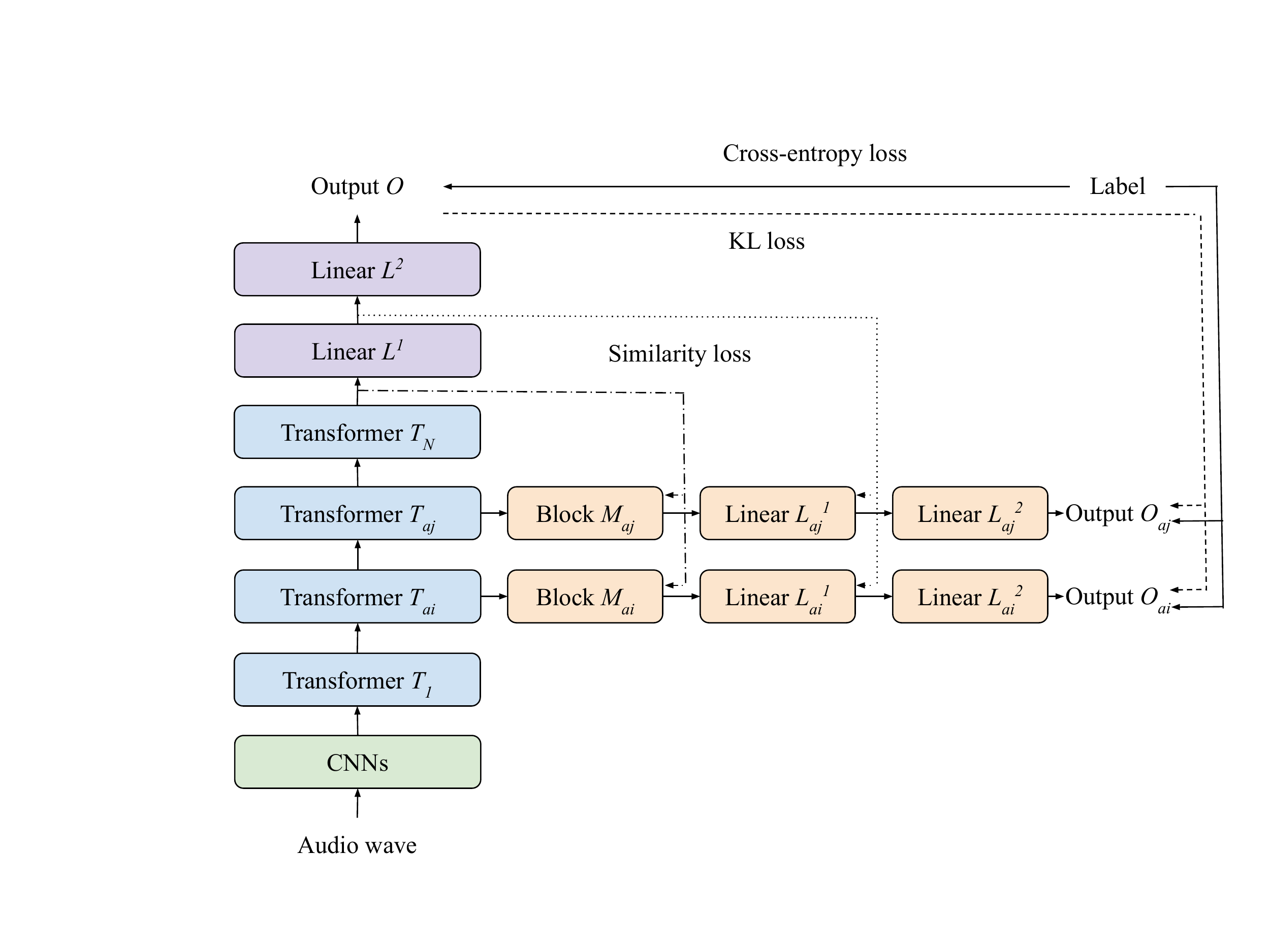}
    \vspace{-15pt}
    \caption{The framework of self-distillation on wav2vec 2.0. The output of an intermediate transformer layer is processed by a block model and two linear layers. The output of each second linear layer is the predicted probabilities on emotional classes. The backward process is implemented via three types of loss functions. `----': cross-entropy loss, `- - -': Kullback–Leibler (KL) divergence loss, `- $\cdot$ - $\cdot$' or `$\cdot\cdot\cdot\cdot\cdot$': similarity loss.}
    \label{fig:framework}
\end{figure}

\sstitle{Teacher Model} We assume the input data of wav2vec 2.0 is represented as $(\bm{X}, y)$, where $\bm{X}$ is the raw speech signals and $y$ is the emotional states. The final (\ie $N$-th) transformer layer $T_N$ of wav2vec 2.0 is followed by two linear layers ($L^1$ and $L^2$) with output dimensions of $D_1$ and $D_2$, where $D_2$ is the number of emotional classes (see Figure~\ref{fig:framework}). Regarding the intermediate features, the output of each transformer layer has a dimension of $(B, F, R)$, where $B$ is the sample number in a batch, $F$ represents the number of time steps, and $R$ denotes the dimension of representations at each time step. With the goal of classification, the $N$-th transformer layer's output is pooled into $H_N$ with a dimension of $(B, R)$ before being fed into the two linear layers.

\sstitle{Student Model} To reduce the model parameters of wav2vec 2.0 with self-distillation, additional layers are added after the intermediate transformer layers of wav2vec 2.0 (see Figure~\ref{fig:framework}). In a student model, the transformer layer $T_{ai}$, $1 \leq ai<N $, is followed by a distillation model, including a block $M_{ai}$ and two linear layers ($L_{ai}^1$ and $L_{ai}^2$). $M_{ai}$ is a neural network, and $L_{ai}^2$ is trained for predicting emotional classes. Herein, as $M_{ai}$ is expected to learn representations similar to those from $T_{N}$, the output of $T_{ai}$ is directly fed into the block model without pooling. Therefore, the output of $M_{ai}$ has a dimension of $(B, F, R)$, and is pooled into $H_{M_{ai}}$ with a dimension of $(B, R)$. $H_{M_{ai}}$ is then fed into $L_{ai}^1$ for further process.

Apart from the above single distillation model, multiple distillation  models could be learnt together in self-distillation to improve the flexibility for different depths of models. For instance, two distillation models in Figure~\ref{fig:framework} are developed after transformer layers $T_{ai}$ and $T_{aj}$ to build SER models with different numbers of model parameters.

% https://docs.google.com/drawings/d/1gmvZZTvrfhRt9r_Dg7LsIdy6D7RuqrcCAfa1FkptEYc/edit
\vspace{-10pt}

\subsubsection{Loss Function}
As the pre-trained wav2vec 2.0 has already strong capability of learning representations from speech, we initialise the teacher model's parameters with the pre-trained wav2vec 2.0.
We assume $N_a$ distillation models are built after transformer layers indexed by $A=\{a1, a2, ...\}$. The representations $H_L$ and $H_{L_{ai}}$ are learnt from $L^1$ and $L_{ai}^1$, respectively. The outputs of the second linear layers $L^2$ and $L_{ai}^2$ are represented as $O$ and $O_{ai}$. The model parameters are optimised with the loss function in self-distillation:
\begin{equation}
    \mathcal{L} = \mathcal{L}_c + \alpha \mathcal{L}_k + \beta \mathcal{L}_s,
    \label{eq:loss}
\end{equation}
where $\mathcal{L}_c$ is the cross-entropy loss, $\mathcal{L}_k$ is the Kullback–Leibler (KL) divergence loss, $\mathcal{L}_s$ is the similarity loss, and $\alpha$ and $\beta$ are constant values.

\sstitle{Cross-entropy Loss}
To train wav2vec 2.0 and distillation models for performing SER, the cross-entropy loss contains i) the cross entropy loss on the teacher model (\ie $L^1$, $L^2$, and wav2vec 2.0), and ii) the cross entropy loss on the student models (\ie distillation models and partial model parameters of wav2vec 2.0):
\begin{equation}
    \mathcal{L}_c=\mathcal{L}_{\mbox{ce}} (O, y) + \gamma \frac{1}{N_a}\sum_{ai \in A} \mathcal{L}_{\mbox{ce}} (O_{ai}, y),
\end{equation}
where $\mathcal{L}_{\mbox{ce}}$ is the typical cross entropy loss for training a model in supervised learning, and $\gamma$ is a constant value.

\sstitle{KL Loss}
The outputs of the distillation models are expected to be similar to that of the linear layer $L^2$, which is the final layer of the teacher model. With this target, the Kullback-Leibler (KL) loss aims to regularise the outputs $O$ and $O_{ai}$:
\begin{equation}
    \mathcal{L}_k=\frac{1}{N_a}\sum_{ai \in A} O \log \frac{O}{O_{ai}}.
\end{equation}

\sstitle{Similarity Loss}
Apart from the loss functions computed on the model outputs, loss functions on the interval features learnt from the intermediate layers can further help train strong student models. 
In this work, we compare three loss functions, including $L_1$, $L_2$, and cosine similarity. Their combinations are also compared with single functions. Furthermore, these loss functions could be either on the outputs of the $N$-th transformer layer and the blocks ($H_N$ and $H_{M_{ai}}$), or on the output of the first linear layers ($H_L$ and $H_{L_{ai}}$):
\begin{equation}
    \mathcal{L}_s=\frac{1}{N_a}\sum_{ai \in A} \mathcal{L}_{\mbox{sim}} (H_N, H_{M_{ai}}) \;  
    \mbox{OR}\; \frac{1}{N_a}\sum_{ai \in A} \mathcal{L}_{\mbox{sim}} (H_L, H_{L_{ai}}),
\end{equation}
where $\mathcal{L}_{\mbox{sim}}$ is a loss function of $L_1$, $L_2$ or negative cosine similarity.

%\subsection{Model training}
%\todo{TODO: explain how we train our model. Is it training wav2vec 2.0 from scratch (similar to \cite{zhang2019your}) or fine-tuning or some adaptive training (similar to 
%\cite{zhang2021self})?}

\subsection{Dynamic Inference}

%\todo{Self-distilled student at deeper level is always better? Or we use some dynamic inference~\cite{zhang2021self}?}

Although the deep layers of wav2vec 2.0 can often learn 
%BS: changed high to higher - please check
higher-level reresentations than shallow ones, self-distillation can provide dynamic inference models~\cite{zhang2021self} via training both shallow and deep student models with good performance. Shallow distillation models require less parameters than deep ones, and deep ones may perform better than shallow ones. The flexibility of self-distillation enables SER applications to be applied to various hardwares, from wearable devices to work stations.

\section{Experiments}
\label{sec:exp}
\subsection{Database}
\label{sec:data}
The database of elicited mood in speech (DEMoS)~\cite{parada2019demos} is used to verify the self-distillation for SER. DEMoS with 9\,365 emotional and 332 neutral Italian speech samples was collected from 68 speakers (f: 23, m: 45)~\cite{parada2019demos}. 
Each speech sample was annotated with one of the eight classes: \emph{anger}, \emph{disgust}, \emph{fear}, \emph{guilt}, \emph{happiness}, \emph{sadness}, \emph{surprise}, and \emph{neutral}.
To implement experiments that can be compared to other studies~\cite{ren2020generating,ren2020enhancing} on DEMoS, the minor \emph{neutral} class is not used, and the speaker-independent training/development/test sets are the same to our prior study~\cite{ren2020generating}. The detail of the data distribution on the seven emotional classes can be found in~\cite{ren2020generating}.

%\todo{TODO: It seems there is no description about the labels.}

\subsection{Experimental Settings}
\label{sec:expsetting}
\sstitle{Evaluations Metrics}
In this study, the unweighted average recall (UAR) is employed to evaluate the performance of SER models. A UAR is computed as the average of all class-wise recalls.

\sstitle{Implementation Details}
With the goal of classifying emotion states, the wav2vec 2.0 model is followed by two linear layers with the numbers of output neurons $\{256, 7\}$, respectively. In each distillation model, the block could be one of the three layers: CNN (number of output channels: $1$, kernel size: $(1, 1)$), Long Short-Time Memory recurrent neural networks (LSTM-RNN) (number of output features: $768$), Gated Recurrent Unit (GRU) RNN (number of output features: $768$). The two linear layers in each distillation model also have the numbers of output neurons $\{256, 7\}$, respectively.

Each model is trained on the training set and validated on the development set, and further trained on the combination of the training and development sets, and validated on the test set. 
During training, the hyperparameters of the loss function are set as $\alpha=\beta=\gamma=1$.
All training procedures of self-distillation are optimised by an Adam optimiser with a learning rate of $3e-5$, and stopped at the $20$-th epoch, when the batch size is $16$. 

\sstitle{Reproducibility Environments}
To improve the reproducibility, the code of this work is released at: \url{https://github.com/leibniz-future-lab/SelfDistill-SER}.
%BS: IMPORTANT - do not foget to add a link here :)

\subsection{Sensitivity Analysis}
\label{sec:sensiticity}
\begin{figure}
    \centering
    \begin{subfigure}[b]{\linewidth}
    \centering
        \includegraphics[width=.7\linewidth]{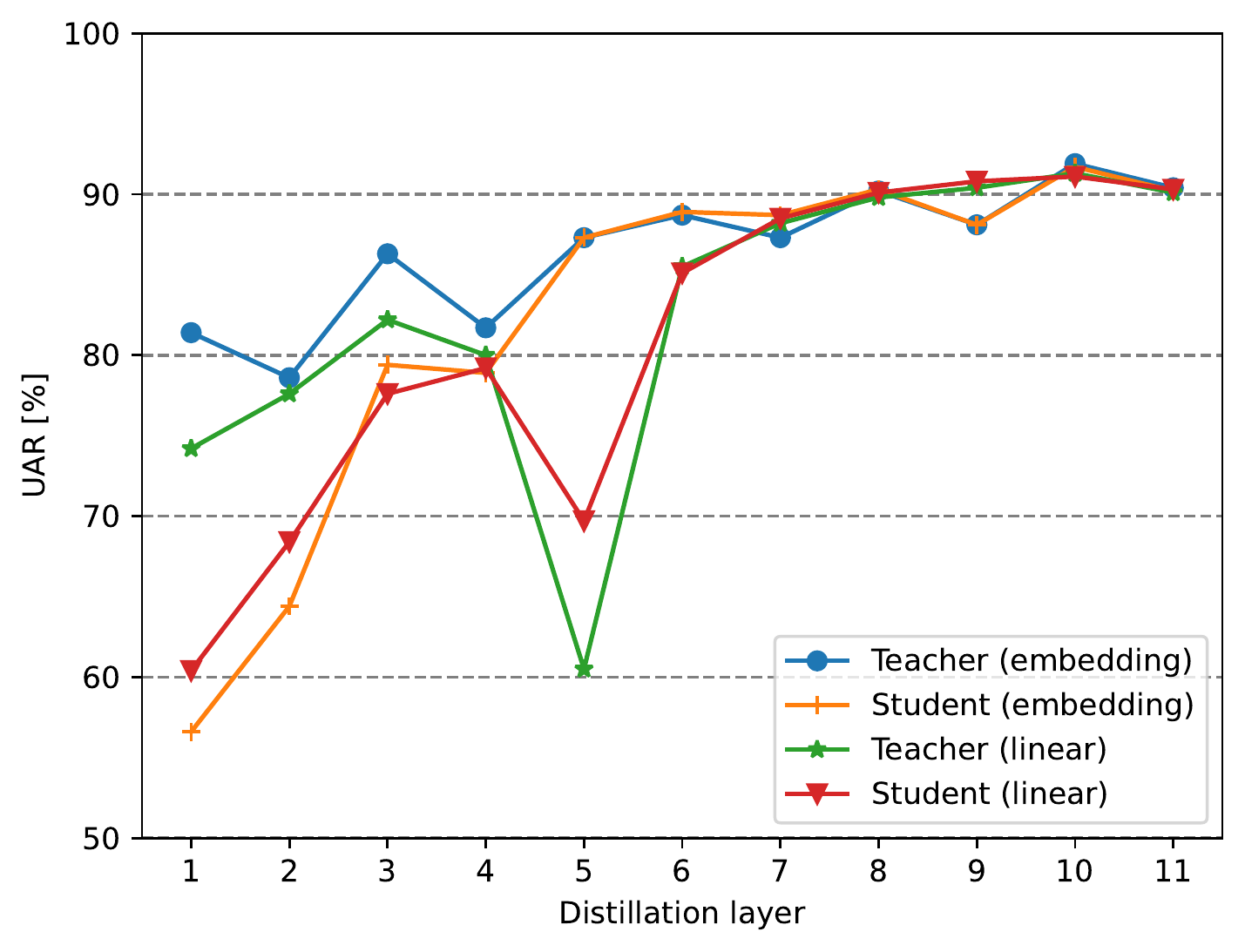}
        \vspace{-10pt}
        \caption{Performance on the development set.}
    \end{subfigure}
    \\
    \begin{subfigure}[b]{\linewidth}
    \centering
        \includegraphics[width=.7\linewidth]{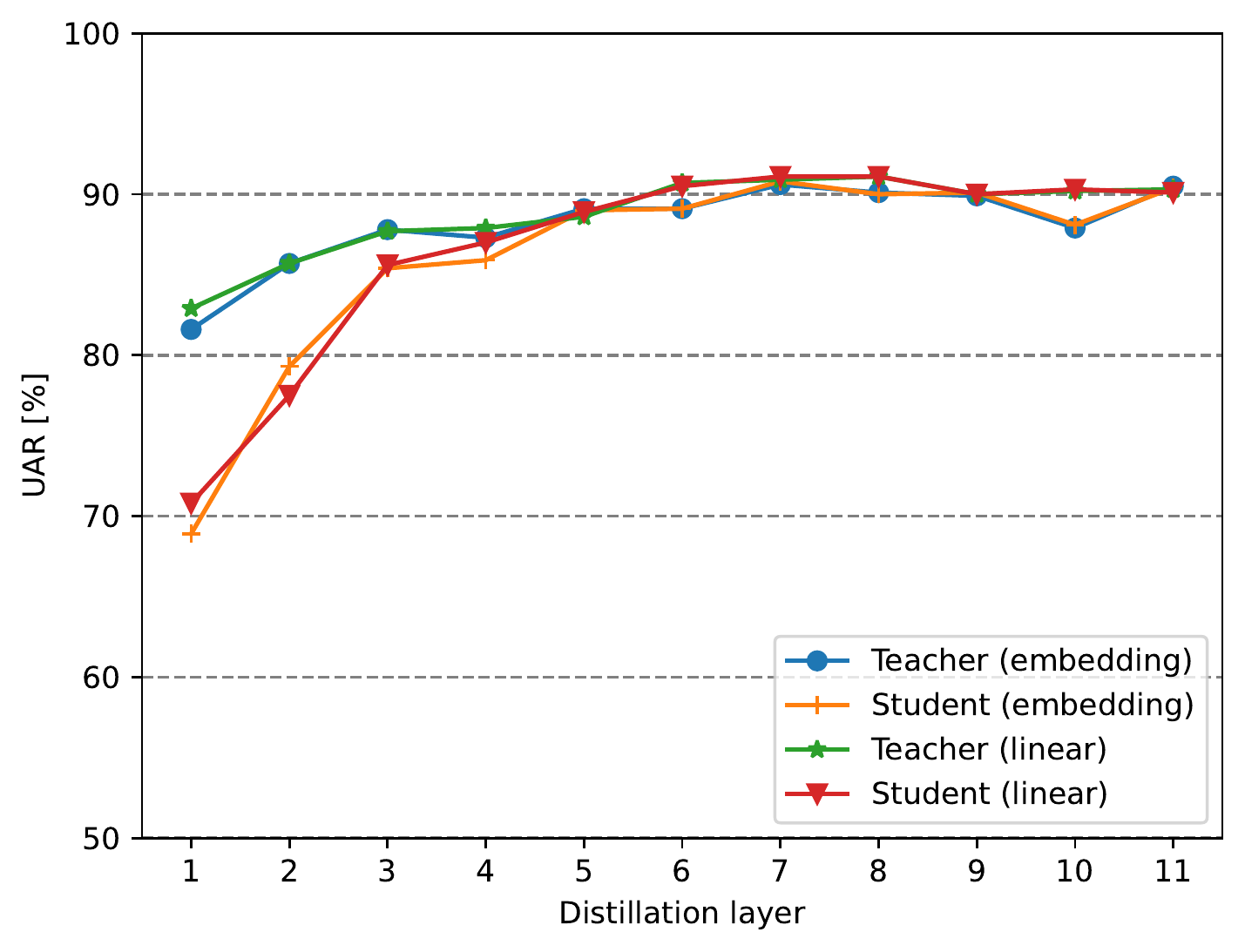}
        \vspace{-10pt}
        \caption{Performance on the test set.}
    \end{subfigure}    
    \vspace{-20pt}
    \caption{Comparison of the performance (UAR [\%]) of the teacher models and the student ones, each of which has single convolutional distillation layer. The loss function is $L_2$ loss.}
    \label{fig:result_singlelayer}
\end{figure}

Figure~\ref{fig:result_singlelayer} shows the performance of self-distillation with single CNN-based distillation model. We can see that, on both development set (Figure~\ref{fig:result_singlelayer} (a)) and test set (Figure~\ref{fig:result_singlelayer} (b)), the performance is increasing when the distillation layer is going deeper. This indicates that deeper model layers of wav2vec 2.0 can learn more abstract representations than shallower ones. The teacher models perform better than shallow student models, but are comparable with deep student models, especially after the $7$-th distillation layer. 

\begin{comment}
\begin{table*}[]
    \caption{Comparison of the performance (UAR [\%]) of self-distillation models, each of which has single convolutional distillation layer. The loss function is $L_2$ loss.}
    \label{tab:sensitivity}
    \centering
    \begin{tabular}{l|c|c|c|c|c|c|c|c|c}
    \toprule
    & \multicolumn{4}{c|}{Embedding} & \multicolumn{4}{c|}{Linear} &  \\
    & \multicolumn{2}{c|}{Deep} & \multicolumn{2}{c|}{Distill} & \multicolumn{2}{c|}{Deep} & \multicolumn{2}{c|}{Distill}  \\
    Distilling layer &  Devel & Test&  Devel & Test &  Devel & Test &  Devel & Test & \#Param \\
    \hline
    1 & 81.4 & 81.6 & 56.6 & 68.9 & 74.2 & 82.9 & 60.4 & 70.8 \\
    2 & 78.6 & 85.7 & 64.4 & 79.3 & 77.6 & 85.7 & 68.4 & 77.5 \\
    3 & 86.3 & 87.8 & 79.4 & 85.4 & 82.2 & 87.7 & 77.6 & 85.6 \\
    4 & 81.7 & 87.3 & 78.9 & 85.9 & 80.0 & 87.9 & 79.2 & 87.0 \\
    \hline
    5 & 87.3 & 89.1 & 87.3 & 89.0 & 60.5 & 88.6 & 69.7 & 88.9 \\
    6 & 88.7 & 89.1 & 88.9 & 89.1 & 85.5 & 90.7 & 85.1 & 90.5 \\
    7 & 87.3 & 90.6 & 88.7 & 90.8 & 88.2 & 90.9 & 88.5 & 91.1 \\
    8 & 90.2 & 90.1 & 90.3 & 90.0 & 89.8 & 91.1 & 90.1 & 91.1 \\
    \hline
    9 & 88.1 & 89.9 & 88.1 & 90.1 & 90.4 & 90.0 & 90.8 & 90.0 \\
    10 & 91.9 & 87.9 & 91.7 & 88.1 & 91.3 & 90.2 & 91.1 & 90.3 \\
    11 & 90.4 & 90.5 & 90.2 & 90.4 & 90.1 & 90.3 & 90.3 & 90.1 \\
    \hline
    Mean & 86.5 & 88.1 & 82.2 & 86.1 & 82.7 & 88.7 & 81.0 & 86.6 \\
    \bottomrule
    \end{tabular}
\end{table*}
\end{comment}
As the distillation layers at the embedding level perform slightly better than those at the linear level, the embedding level is selected in the next experiments for self-distillation. 
To provide different model sizes with self-distillation, we group the distillation layers into three groups and select layers which have the best performance on the development set. Therefore, we use layers \{3, 8, 10\} for multi-layer self-distillation in the following experiments.

% % orginal result on CNN (Table 2, now moved to the following table.)
%\begin{table*}[]
%    \centering
%    \caption{Comparison of the performance (UAR [\%]) of self-distillation with CNN models optimised by different loss functions. The distillation layers \{3, 8, 10\} are selected from Table 1.}
%    \label{tab:sensitivity}
%    \begin{tabular}{l|c|c|c|c|c|c|c|c|c|c|c|c}
%    \toprule
%    & \multicolumn{2}{c}{Deep} & \multicolumn{2}{c}{Layer 1} &\multicolumn{2}{c}{Layer 2} &\multicolumn{2}{c}{Layer 3} &\multicolumn{2}{c}{Fusion}\\
%    Loss &  Devel & Test &  Devel & Test &  Devel & Test&  Devel & Test&  Devel & Test\\
%    \hline
%       $L_1$ & 87.3 & 89.0 & 75.2 & 82.4 & 86.9 & 88.8 & 87.1 & 88.9 & 85.7 & 88.3    \\
%       $L_2$ & 88.3 & 89.3 & 70.2 & 81.6 & 88.6 & 89.4 & 88.2 & 89.2 & 86.8 & 88.5    \\
%       Cosine sim. & 89.6 & 89.7 & 76.0 & 81.5 & 89.5 & 89.9 & 89.7 & 89.7 & 88.7 & 89.8    \\
%       $L_1$ + Cosine sim. & 88.4 & 87.8 & 77.4 & 77.5 & 88.4 & 88.1 & 88.4 & 87.8 & 87.5 & 87.4    \\
%       $L_2$ + Cosine sim. & 88.1 & 89.7 & 77.5 & 81.5 & 88.8 & 89.3 & 88.4 & 89.5 & 87.4 & 89.0    \\
%    \bottomrule
%    \end{tabular}
%\end{table*}

\begin{table*}[htp]
    \caption{Comparison of the performance (UAR [\%]) of multi-layer self-distillation (distillation layers: \{3, 8, 10\}). The best performance of the deepest model is highlighted with `\_\_', and the best performance of distillation layers is highlighted with bold fonts.}
    \vspace{-10pt}
    \label{tab:sensitivity}
    \centering
    \scalebox{0.85}{
    \begin{tabular}{l|l|P{0.6cm}|P{0.6cm}|P{0.6cm}|P{0.6cm}|P{0.6cm}|P{0.6cm}|P{0.6cm}|P{0.6cm}|P{0.6cm}|P{0.6cm}}
    \toprule
    & &\multicolumn{2}{c|}{\textbf{Deepest}} & \multicolumn{2}{c|}{\textbf{Layer 3}} & \multicolumn{2}{c|}{\textbf{Layer 8}} & \multicolumn{2}{c|}{\textbf{Layer 10}} & \multicolumn{2}{c}{\textbf{Fusion}} \\
    \cline{3-12}
    \emph{NN} & \emph{Loss}& \emph{Devel} & \emph{Test} & \emph{Devel} & \emph{Test} & \emph{Devel} & \emph{Test} & \emph{Devel} & \emph{Test} & \emph{Devel} & \emph{Test} \\
    %Fine-tuning(epoch15) &--&89.7&90.9  \\
    \hline
    &$L_1$ & 87.3 & 89.0 & 75.2 & 82.4 & 86.9 & 88.8 & 87.1 & 88.9 & 85.7 & 88.3    \\
    &$L_2$ & 88.3 & 89.3 & 70.2 & 81.6 & 88.6 & 89.4 & 88.2 & 89.2 & 86.8 & 88.5    \\
    Self-distillation w/ CNN 
    & Cosine sim. & \underline{89.6} & \underline{89.7} & \textbf{76.0} & \textbf{81.5} & \textbf{89.5} & \textbf{89.9} & \textbf{89.7} & \textbf{89.7} & \textbf{88.7} & \textbf{89.8}    \\
    &$L_1$ + Cosine sim. & 88.4 & 87.8 & 77.4 & 77.5 & 88.4 & 88.1 & 88.4 & 87.8 & 87.5 & 87.4    \\
    &$L_2$ + Cosine sim. & 88.1 & 89.7 & 77.5 & 81.5 & 88.8 & 89.3 & 88.4 & 89.5 & 87.4 & 89.0    \\
    \hline
    &$L_1$ & 91.4 & 90.2 & \textbf{83.3} & \textbf{85.7} & \textbf{91.7} & \textbf{90.7} & \textbf{91.8} & \textbf{90.5} & \textbf{90.5} & \textbf{89.9} \\
    &$L_2$ & 90.7 & 91.2 & 78.5 & 86.7 & 91.0 & 91.1 & 90.8 & 90.9 & 89.6 & 90.8 \\
    Self-distillation w/ LSTM
    & Cosine sim. & 90.8 & 88.8 & 79.2 & 75.2 & 91.1 & 87.9 & 90.8 & 88.5 & 90.1 & 87.1    \\
    &$L_1$ + Cosine sim. & 90.3 & 90.6 & 77.7 & 86.5 & 90.1 & 90.3 & 90.3 & 90.5 & 88.8 & 90.5 \\
    &$L_2$ + Cosine sim. & \underline{91.8} & \underline{91.4} & 83.3 & 84.7 & 91.1 & 91.4 & 91.7 & 91.2 & 90.1 & 90.9  \\
    \hline
    &$L_1$ & 86.3 & 91.0 & 80.2 & 85.5 & 84.7 & 91.6 & 84.5 & 91.2 & 86.1 & 90.8   \\
    &$L_2$ & 90.9 & 90.7 & \textbf{82.0} & \textbf{84.1} & \textbf{91.2} & \textbf{90.4} & \textbf{91.1} & \textbf{90.6} & \textbf{90.0} & \textbf{90.0}  \\
    Self-distillation w/ GRU 
    & Cosine sim. & 90.2 & 89.6 & 74.8 & 82.2 & 90.3 & 89.5 & 89.9 & 89.4 & 88.7 & 88.9    \\
    &$L_1$ + Cosine sim. & \underline{91.2} & \underline{90.0} & 79.5 & 83.7 & 91.4 & 90.2 & 91.0 & 90.1 & 89.5 & 89.7  \\
    &$L_2$ + Cosine sim. & 88.1 & 91.6 & 72.2 & 85.1 & 87.5 & 91.5 & 86.5 & 91.7 & 86.8 & 90.7  \\
    \bottomrule
    \end{tabular}
    }
\end{table*}

\subsection{Ablation Study}
\label{sec:ablation}
%On the average UARs over the three distillation layers on the development set, cosine similarity performs best among the five loss functions. So we use cosine similarity in the next experiments.

With the distillation layers \{3, 8, 10\}, we compare the three block models (\ie CNN, LSTM-RNN, and GRU-RNN) with various similarity loss functions (\ie $\mathcal{L}_{\mbox{sim}}$). From Table~\ref{tab:sensitivity}, we can find that all similarity loss functions perform comparably for each block model. Particularly, the single similarity loss functions (\ie $L_1$, $L2$, and cosine similarity) perform better than the combinations of them. This might be related to the setting of hyperparameters in the loss functions. Furthermore, the LSTM-RNN and the GRU-RNN models outperform the CNN one when comparing the three models blocks. This may be because RNNs can better learn sequential information than CNNs. Regarding the self-distillation, the performance of the student models at layers 8 and 10 is comparable with the deepest model. Layer 3 performs slightly worse than layers 8 and 10, as the corresponding student model of layer 3 is shallower than those of layers 8 and 10. Finally, the fusion results of the three distillation models are comparable with the results of layer 10. 

\subsection{Comparison with SOTA}
\label{sec:compare}
We compared the results of self distillation with the other SOTA models, including the following three groups of models. 
(1) CNN-4, VGG-16, ResNet-50, and VGG-16 with adversarial training are trained from scratch~\cite{ren2020generating}. 
(2) The models of wav2vec 2.0 with fine-tuning are trained for 20 epochs based on the pre-trained wav2vec 2.0, when the transformer layers after finetuned layers are frozen. 
(3) The layer-wise models are trained via the layer-wise knowledge distillation in~\cite{chang2022distilhubert}. As wav2vec 2.0 has 12 transformer layers, which is the same as the number of encoder layers in HuBERT in~\cite{chang2022distilhubert}, the layer-wise models are mostly developed according to the settings in ~\cite{chang2022distilhubert}. The teacher model is the pre-trained wav2vec 2.0, and the student model is part of the pre-trained wav2vec 2.0 itself (from the first layer to the second transformer layer). Notably, all parameters of the teacher model are frozen. The layer-wise models are trained at two stages: i) training the student model to predict layers \{4, 8, 12\}) of the teacher model, and ii) fine-tuning the student model on DEMoS for SER. To train a strong student model, the first stage is trained with 20 epochs. To implement fair experimental comparisons, the second stage also consists of 20 epochs. 

As wav2vec 2.0 was pre-trained on the large-scale speech database, The models based on wav2vec 2.0 are mostly better than models trained from scratch (CNN-4, VGG-16 (+ adversarial training), and ResNet-50). When comparing fine-tuned wav2vec 2.0 models and self-distillation, the student model at layer 3 outperforms the corresponding fine-tuned one. The fine-tuned models are comparable with self-distillation at layers 8 and 10, while self-distillation trains student models at different layers in single training procedure. Our self-distillation outperforms layer-wise distillation at all three layers. This may be caused by the shallower student models (encoder and two transformer layers) in layer-wise distillation. For a specific task, self-distillation requires less training epochs ($20$ in our work) than layer-wise distillation (40 in our study), increasing the training efficiency.

%BS: I added an hline and extended the caption to highlight which are our approaches - please check.
\begin{table}[]
    \caption{Comparison of the performance (UAR [\%]) between our approach (lower lines) and the state-of-the-art (SOTA).}
    \vspace{-10pt}
    \label{tab:sensitivity}
    \centering
    \scalebox{0.85}{
    \begin{tabular}{l|P{0.8cm}|P{0.8cm}|R{1.2cm}}
    \toprule
    \textbf{NN} &  \textbf{Devel} & \textbf{Test} & \textbf{\#Param}\\
    \hline
      CNN-4~\cite{ren2020generating} & 82.6 & 83.6 & 4.3\,M\\
      VGG-16~\cite{ren2020generating}  & 79.8 & 83.6 & 14.7\,M\\
      ResNet-50~\cite{ren2020generating}  & 71.9& 81.3 & 23.5\,M\\
      VGG-16 + adversarial training~\cite{ren2020generating} & 87.5 & 86.7 & 14.7\,M\\
      Wav2vec2 (layer 3) + fine-tuning & 77.1 & 83.4 & 31.5\,M \\
      Wav2vec2 (layer 8) + fine-tuning & 90.1 & 90.9 & 66.9\,M \\
      Wav2vec2 (layer 10) + fine-tuning &91.7 & 89.2 & 81.1\,M \\
      Wav2vec2 (deepest) + fine-tuning & 91.1 & 90.6 & 95.2\,M \\ % 95243271        
      Layer-wise distillation w/ CNN  & 54.3 & 72.5 & 24.4\,M \\ 
      Layer-wise distillation w/ LSTM & 70.8 & 79.0 & 38.5\,M \\  
      Layer-wise distillation w/ GRU & 73.1 & 77.4 & 35.0\,M\\  
      \hline
      \textbf{Self-distillation (layer 3)} & 83.3 & 85.7 & 36.2\,M\\%36178695
      \textbf{Self-distillation (layer 8)} & 91.7 & 90.7 & 71.6\,M\\ %71618055 
      \textbf{Self-distillation (layer 10)} & 91.8 & 90.5 & 85.8\,M \\%85793799 
      \textbf{Self-distillation (teacher)} & \textbf{91.8} & \textbf{91.4} & 100.0\,M\\ %99969543      
    \bottomrule
    \end{tabular}
    }
\end{table}

\section{Conclusions and Future Work}
\label{sec:conclusion}
%BS: (only) outlook in past tense - I changed
This work aimed to reduce model parameters via self-distillation for fast and effective speech emotion recognition. The experiments were implemented on the Database of Elicited Mood in Speech (DEMoS)~\cite{parada2019demos} with the pre-trained wav2vec 2.0. The experimental results demonstrated that the student model at a shallow layer (layer 3) outperformed the corresponding fine-tuned wav2vec 2.0, and self-distillation achieved comparable performance with that of fine-tuned wav2vec 2.0 at deep layers. Moreover, self-distillation performed better than layer-wise knowledge distillation. 
In future work, the self-distillation approach will be verified on multiple databases. We will also investigate to further reduce the wav2vec 2.0 model by the state-of-the-art model compression approaches~\cite{berthelier2021deep,tan2021towards}.

\vfill\pagebreak
\balance
% References should be produced using the bibtex program from suitable
% BiBTeX files (here: strings, refs, manuals). The IEEEbib.bst bibliography
% style file from IEEE produces unsorted bibliography list.
% -------------------------------------------------------------------------
\bibliographystyle{IEEEbib}
\bibliography{refs}

\begin{thebibliography}{10}

\bibitem{schuller2018speech}
Bj{\"o}rn~W Schuller,
\newblock ``Speech emotion recognition: Two decades in a nutshell, benchmarks,
  and ongoing trends,''
\newblock {\em Communications of the ACM}, vol. 61, no. 5, pp. 90--99, 2018.

\bibitem{el2011survey}
Moataz El~Ayadi, Mohamed~S Kamel, and Fakhri Karray,
\newblock ``Survey on speech emotion recognition: Features, classification
  schemes, and databases,''
\newblock {\em Pattern recognition}, vol. 44, no. 3, pp. 572--587, 2011.

\bibitem{khalil2019speech}
Ruhul~Amin Khalil, Edward Jones, Mohammad~Inayatullah Babar, Tariqullah Jan,
  Mohammad~Haseeb Zafar, and Thamer Alhussain,
\newblock ``Speech emotion recognition using deep learning techniques: A
  review,''
\newblock {\em IEEE Access}, vol. 7, pp. 117327--117345, 2019.

\bibitem{baevski2020wav2vec}
Alexei Baevski, Yuhao Zhou, Abdelrahman Mohamed, and Michael Auli,
\newblock ``wav2vec 2.0: A framework for self-supervised learning of speech
  representations,''
\newblock in {\em Proc.\ NeurIPS}, Vancouver, Canada, 2020, pp. 1--12.

\bibitem{SchneiderBCA19}
Steffen Schneider, Alexei Baevski, Ronan Collobert, and Michael Auli,
\newblock ``wav2vec: Unsupervised pre-training for speech recognition,''
\newblock in {\em Proc.\ INTERSPEECH}, Graz, Austria, 2019, pp. 3465--3469.

\bibitem{BaevskiSA20}
Alexei Baevski, Steffen Schneider, and Michael Auli,
\newblock ``vq-wav2vec: Self-supervised learning of discrete speech
  representations,''
\newblock in {\em ICLR}, Virtual, 2020.

\bibitem{chang2022distilhubert}
Heng-Jui Chang, Shu-wen Yang, and Hung-yi Lee,
\newblock ``Distilhubert: Speech representation learning by layer-wise
  distillation of hidden-unit bert,''
\newblock in {\em Proc.\ ICASSP}, Singapore, 2022, pp. 7087--7091.

\bibitem{zhang2019your}
Linfeng Zhang, Jiebo Song, Anni Gao, Jingwei Chen, Chenglong Bao, and Kaisheng
  Ma,
\newblock ``Be your own teacher: Improve the performance of convolutional
  neural networks via self distillation,''
\newblock in {\em Proc.\ ICCV}, Seoul, Korea, 2019, pp. 3713--3722.

\bibitem{zhang2021self}
Linfeng Zhang, Chenglong Bao, and Kaisheng Ma,
\newblock ``Self-distillation: Towards efficient and compact neural networks,''
\newblock {\em IEEE Transactions on Pattern Analysis and Machine Intelligence},
  vol. 44, no. 8, pp. 4388--4403, 2021.

\bibitem{gong2022ssast}
Yuan Gong, Cheng-I Lai, Yu-An Chung, and James Glass,
\newblock ``Ssast: Self-supervised audio spectrogram transformer,''
\newblock in {\em Proc.\ AAAI}, Virtual, 2022, vol.~36, pp. 10699--10709.

\bibitem{singh2021multimodal}
Prabhav Singh, Ridam Srivastava, KPS Rana, and Vineet Kumar,
\newblock ``A multimodal hierarchical approach to speech emotion recognition
  from audio and text,''
\newblock {\em Knowledge-Based Systems}, vol. 229, pp. 107316, 2021.

\bibitem{wani2021comprehensive}
Taiba~Majid Wani, Teddy~Surya Gunawan, Syed Asif~Ahmad Qadri, Mira Kartiwi, and
  Eliathamby Ambikairajah,
\newblock ``A comprehensive review of speech emotion recognition systems,''
\newblock {\em IEEE Access}, vol. 9, pp. 47795--47814, 2021.

\bibitem{pepino2021emotion}
Leonardo Pepino, Pablo Riera, and Luciana Ferrer,
\newblock ``Emotion recognition from speech using wav2vec 2.0 embeddings,''
\newblock in {\em Proc.\ INTERSPEECH}, Brno, Czech Republic, 2021, pp.
  3400--3404.

\bibitem{chen2021exploring}
Li-Wei Chen and Alexander Rudnicky,
\newblock ``Exploring wav2vec 2.0 fine-tuning for improved speech emotion
  recognition,''
\newblock {\em arXiv preprint arXiv:2110.06309}, 2021.

\bibitem{wagner2022dawn}
Johannes Wagner, Andreas Triantafyllopoulos, Hagen Wierstorf, Maximilian
  Schmitt, Florian Eyben, and Bj{\"o}rn~W Schuller,
\newblock ``Dawn of the transformer era in speech emotion recognition: closing
  the valence gap,''
\newblock {\em arXiv preprint arXiv:2203.07378}, 2022.

\bibitem{mobahi2020self}
Hossein Mobahi, Mehrdad Farajtabar, and Peter Bartlett,
\newblock ``Self-distillation amplifies regularization in hilbert space,''
\newblock in {\em Proc.\ NeurIPS}, Vancouver, Canada, 2020, pp. 1--11.

\bibitem{pham2022revisiting}
Minh Pham, Minsu Cho, Ameya Joshi, and Chinmay Hegde,
\newblock ``Revisiting self-distillation,''
\newblock {\em arXiv preprint arXiv:2206.08491}, 2022.

\bibitem{lee2020self}
Hankook Lee, Sung~Ju Hwang, and Jinwoo Shin,
\newblock ``Self-supervised label augmentation via input transformations,''
\newblock in {\em Proc.\ ICML}, Virtual, 2020, pp. 5714--5724.

\bibitem{vaessen2022fine}
Nik Vaessen and David~A Van~Leeuwen,
\newblock ``Fine-tuning wav2vec2 for speaker recognition,''
\newblock in {\em Proc.\ ICASSP}, Singapore, 2022, pp. 7967--7971.

\bibitem{mohamed2021arabic}
Omar Mohamed and Salah~A Aly,
\newblock ``Arabic speech emotion recognition employing wav2vec2. 0 and hubert
  based on baved dataset,''
\newblock {\em arXiv preprint arXiv:2110.04425}, 2021.

\bibitem{panayotov2015librispeech}
Vassil Panayotov, Guoguo Chen, Daniel Povey, and Sanjeev Khudanpur,
\newblock ``{Librispeech: An ASR corpus based on public domain audio books},''
\newblock in {\em Proc.\ ICASSP}, Brisbane, Australia, 2015, pp. 5206--5210.

\bibitem{wongpatikaseree2022real}
Konlakorn Wongpatikaseree, Sattaya Singkul, Narit Hnoohom, and Sumeth Yuenyong,
\newblock ``Real-time end-to-end speech emotion recognition with cross-domain
  adaptation,''
\newblock {\em Big Data and Cognitive Computing}, vol. 6, no. 3, pp. 79, 2022.

\bibitem{parada2019demos}
Emilia Parada-Cabaleiro, Giovanni Costantini, Anton Batliner, Maximilian
  Schmitt, and Bj{\"o}rn Schuller,
\newblock ``{DEMoS: An Italian emotional speech corpus},''
\newblock {\em Language Resources and Evaluation}, pp. 1--43, Feb. 2019.

\bibitem{ren2020generating}
Zhao Ren, Alice Baird, Jing Han, Zixing Zhang, and Bj{\"o}rn Schuller,
\newblock ``{Generating and protecting against adversarial attacks for deep
  speech-based emotion recognition models},''
\newblock in {\em Proc.\ ICASSP}, Barcelona, Spain, 2020, pp. 7184--7188.

\bibitem{ren2020enhancing}
Zhao Ren, Jing Han, Nicholas Cummins, and Bj{\"o}rn Schuller,
\newblock ``{Enhancing transferability of black-box adversarial attacks via
  lifelong learning for speech emotion recognition models},''
\newblock in {\em Proc.\ INTERSPEECH}, Shanghai, China, 2020, pp. 496--500.

\bibitem{berthelier2021deep}
Anthony Berthelier, Thierry Chateau, Stefan Duffner, Christophe Garcia, and
  Christophe Blanc,
\newblock ``Deep model compression and architecture optimization for embedded
  systems: A survey,''
\newblock {\em Journal of Signal Processing Systems}, vol. 93, no. 8, pp.
  863--878, 2021.

\bibitem{tan2021towards}
Ke~Tan and DeLiang Wang,
\newblock ``Towards model compression for deep learning based speech
  enhancement,''
\newblock {\em IEEE/ACM transactions on audio, speech, and language
  processing}, vol. 29, pp. 1785--1794, 2021.

\end{thebibliography}

\end{document}